\documentstyle[12pt]{article}
      \catcode`\ä=\active \defä{\"a}
      \catcode`\ö=\active \defö{\"o}
      \catcode`\ü=\active \defü{\"u}
      \catcode`\Ä=\active \defÄ{\"A}
      \catcode`\Ö=\active \defÖ{\"O}
      \catcode`\Ü=\active \defÜ{\"U}
      \catcode`\ß=\active \defß{{\ss}} 
\def\format#1#2#3#4{
  \textwidth=#3
  \textheight=#4
  \topmargin=#2
  \advance\topmargin by -1in
  \evensidemargin=#1
  \oddsidemargin=#1
  \advance\evensidemargin by -1in
  \advance\oddsidemargin by -1in
}
\def\nohead{
\headheight=0pt
\headsep=0pt
\topskip=0pt
}
\nohead\format{27.5mm}{24mm}{155mm}{240mm}
\newlength{\displaystylematrixsep}
\def\protozahlen#1#2#3#4{{%
\mathord{\hbox{$#1\rm #2\hskip -#3 #4$}}%
}}

\def\nz{{\mathchoice{%
\protozahlen{\displaystyle}{I}{2pt}{N}}{%
\protozahlen{\displaystyle}{I}{2pt}{N}}{%
\protozahlen{\scriptstyle}{I}{1.4pt}{N}}{%
\protozahlen{\scriptscriptstyle}{I}{1.2pt}{N}%
}}}

\def\rz{{\mathchoice{%
\protozahlen{\displaystyle}{I}{2pt}{R}}{%
\protozahlen{\displaystyle}{I}{2pt}{R}}{%
\protozahlen{\scriptstyle}{I}{1.4pt}{R}}{%
\protozahlen{\scriptscriptstyle}{I}{1.2pt}{R}%
}}}

\def\E{{\mathchoice{%
\protozahlen{\displaystyle}{I}{2pt}{E}}{%
\protozahlen{\displaystyle}{I}{2pt}{E}}{%
\protozahlen{\scriptstyle}{I}{1.4pt}{E}}{%
\protozahlen{\scriptscriptstyle}{I}{1.2pt}{E}%
}}}

\def\defeq{{\mathchoice{%
\mathrel{\mskip\thickmuskip\raise.35pt\hbox{$\mathord{\displaystyle:}$}%
\hbox{$\mathord{\displaystyle=}$}\mskip\thickmuskip}}{%
\mathrel{\mskip\thickmuskip\raise.35pt\hbox{$\mathord{\displaystyle:}$}%
\hbox{$\mathord{\displaystyle=}$}\mskip\thickmuskip}}{%
\mathrel{\mskip.25\thinmuskip\raise.25pt\hbox{$\mathord{\scriptstyle:}$}%
\hbox{$\mathord{\scriptstyle=}$}\mskip.25\thinmuskip}}{%
\mathrel{\mskip.1\thinmuskip\raise.1pt\hbox{$\mathord{\scriptscriptstyle:}$}%
\hbox{$\mathord{\scriptscriptstyle=}$}\mskip.1\thinmuskip}%
}}}
\def\eqdef{{\mathchoice{%
\mathrel{\mskip\thickmuskip\hbox{$\mathord{\displaystyle=}$}%
\raise.35pt\hbox{$\mathord{\displaystyle:}$}\mskip\thickmuskip}}{%
\mathrel{\mskip\thickmuskip\hbox{$\mathord{\displaystyle=}$}%
\raise.35pt\hbox{$\mathord{\displaystyle:}$}\mskip\thickmuskip}}{%
\mathrel{\mskip.25\thinmuskip\hbox{$\mathord{\scriptstyle=}$}%
\raise.25pt\hbox{$\mathord{\scriptstyle:}$}\mskip.25\thinmuskip}}{%
\mathrel{\mskip.1\thinmuskip\hbox{$\mathord{\scriptscriptstyle=}$}%
\raise.1pt\hbox{$\mathord{\scriptscriptstyle:}$}\mskip.1\thinmuskip}%
}}}
\def\halmos{{%
\hspace*{\fill}\hbox to 18pt {\hfill\vrule width 9pt height 9pt depth 0pt}%
}\par\ignorespaces}
\def\e{{\rm e}}
\def\d{{\rm d}}

\def\tr{\mathop {\rm Tr}\nolimits }

\def\esssup{\mathop{\hbox{ess$\,$sup}}}
\def\L{{\rm L}}

\let\eps=\varepsilon
\let\rho=\varrho

\def\vgl{{\frenchspacing \vgl\ }}
\def\bzw{{\frenchspacing \bzw\ }}
\def\sog{{\frenchspacing \sog\ }}

\let\Ds=\displaystyle
\let\bar=\overline

\catcode`\@=11
\def\empty{}
\def\be#1{\def\remembereqnoi{#1}\begin{equation}}
\def\ee{\hbox{\label{\remembereqnoi}}\end{equation}}
\def\btwoe#1#2{%
 \def\remembereqnoi{#1}\def\remembereqnoii{#2}
 \def\neuezeile{\hbox{\label{\remembereqnoi}}\\}
 \begin{eqnarray}
}
\def\etwoe{%
 \hbox{\label{\remembereqnoii}}\end{eqnarray}
}
\def\widecr{\cr\noalign{\vskip\displaystylematrixsep}}
\@addtoreset{equation}{section}

\def\sectionname{}
\def\appendixname{Appendix}
\def\appendix{\par
  \setcounter{section}{0}
  \setcounter{subsection}{0}
  \def\thesection{}
  \let\sectionname\appendixname
  }
\def\abschnitt#1#2{
\refstepcounter{section}\label{#1}
\advance\c@section by -1
\let\oldthesection\thesection
\ifx\sectionname\empty\relax%
\else\gdef\thesection{\sectionname~\oldthesection}%
\fi%
\section{\protect\raggedright #2}%
\let\thesection\oldthesection%
}

\def\xref#1{(\ref{#1})}
\edef\kommaspace{\egroup, \hbox\bgroup}
\edef\komma{,}
\catcode`\,=\active
\let,=\komma
\def\zit#1#2{{\glossary{#2}%
\def\test{#1}%
\ifx\test\empty%
[{\let,=\kommaspace \hbox\bgroup #2\egroup}]%
\else%
[{\let,=\kommaspace \hbox\bgroup #2\egroup}, {\frenchspacing #1}]%
\fi{}%
}}
\newcounter{structure}[section]
\def\thestructure{\thesection.\arabic{structure}}
\newcounter{substructure}[structure]

\def\preknoedel#1{%
\refstepcounter{substructure}\label{\structnummer-#1)}%
\item}
\def\einruecken{%
\def\finishii{\end{list}}%
\begin{list}{}{\leftmargin1.5em \rightmargin0pt \topsep0pt}\item%
}
\newenvironment{structure}[2]{%
\def\structnummer{#1}%
\let\finishi=\relax%
\let\finishii=\relax%
\def\knoedel##1{%
\def\finishi{\end{list}}%
\begin{list}{{\it\roman{substructure})\/}}{%
\leftmargin24pt \labelsep8pt \labelwidth16pt \itemindent0pt%
\rightmargin0pt \topsep0pt}%
\let\knoedel=\preknoedel%
\knoedel{##1}%
}%
\refstepcounter{structure}\label{\structnummer}%
\subsubsection*{#2 \protect\thestructure}%
}{\finishi\finishii\vskip\baselineskip}
\catcode`\@=12
     \def\ssp{U}
\begin{document}
%-----------------------------------------------------------------------------
     \pagestyle{empty}
%-----------------------------------------------------------------------------
\vspace*{20mm}
\begin{center}
\Large\bf
   The Fate of Lifshitz Tails in Magnetic Fields\\[10mm]
\large
   Kurt Broderix$^{1)}$, Dirk Hundertmark$^{2)}$,\\
   Werner Kirsch$^{2)}$, Hajo Leschke$^{3)}$\\[10mm]
\normalsize
\begin{minipage}{105mm}
\begin{itemize}
   \item[$^{1)}$]
      Institut f\"ur Theoretische Physik\\
      Universit\"at G\"ottingen\\
      Bunsenstr. 9, D-37073 G\"ottingen, Germany
   \item[$^{2)}$]
      Institut f\"ur Mathematik\\
      Ruhr-Universit\"at Bochum\\
      Universit\"atsstr. 150, D-44780 Bochum, Germany
   \item[$^{3)}$]
      Institut f\"ur Theoretische Physik\\
      Universit\"at Erlangen-N\"urnberg\\
      Staudtstr. 7, D-91058 Erlangen, Germany
\end{itemize}
\end{minipage}
\end{center}
\vskip20mm

\begin{center}
\begin{minipage}{145mm}
\subsubsection*{Abstract}
We investigate the integrated density of states of the Schr\"odinger
operator in the Euclidean plane with a perpendicular constant magnetic
field and a random potential. For a Poisson random potential with a
non-negative algebraically decaying single-impurity potential we prove
that the leading asymptotic behaviour for small energies is always
given by the corresponding classical result in contrast to the case of
vanishing magnetic field. We also show that the integrated density of
states of the operator restricted to the eigenspace of any Landau
level exhibits the same behaviour. For the lowest Landau level, this
is in sharp contrast to the case of a Poisson random potential with a
delta-function impurity potential.\\[14pt] {\bf Key words:} 
Random Schr\"odinger operators, magnetic fields, Lifshitz tails
\end{minipage}
\end{center}
%-----------------------------------------------------------------------------
     \vskip10mm\par$$\hbox{\leaders\hrule height 1pt\hskip90mm}$$\vskip6mm\par\centerline{Version of February, 22 1995}\par\centerline{To appear in Journal of Statistical Physics}\par
     \newpage
%-----------------------------------------------------------------------------
     \pagestyle{plain}
%-----------------------------------------------------------------------------
\abschnitt{I}{Introduction}
%-----------------------------------------------------------------------------
Random Schr\"odinger operators are differential operators on
$\L^2(\rz^d)$ formally given by $-\frac{1}{2}\nabla^2 + V_\omega$,
where $V_\omega$ is an ergodic (or metrically transitive) random
scalar potential. Here $\nabla$ denotes the nabla operator in the
$d$-dimensional Euclidean space $\rz^d$ and $\L^2(\rz^d)$ is the
Hilbert space of Lebesgue square-integrable complex-valued functions
on $\rz^d$. These operators have been thoroughly investigated by
physicists as well as mathematicians. See \cite{pafi92}, \cite{cala90}
or \cite{kir89} for reviews of basic concepts and rigorous results. For
a more physical point of view see \cite{shef84}, \cite{ligr88}. The
generalization of random Schr\"odinger operators to non-zero constant
magnetic field are operators of the form
\be{2}
   H(V_\omega) \defeq \frac{1}{2}( i \nabla + A)^2 + V_\omega , 
\ee 
where $A:\rz^d\to\rz^d $ is a non-random vector potential such
that the magnetic-field tensor ($B_{jk}$) given by $B_{jk} \defeq
\frac{\partial A_j}{\partial x_k} -\frac{\partial A_k}{\partial x_j}$
is constant. The two-dimensional version of \xref{2} is widely
believed to serve as a minimal model for the integer quantum Hall
effect \cite{sto92}, \cite{javi94}, \cite{beel94} and has therefore
been intensively investigated by physicists, see for instance
\cite{anfo82}, \cite{kume88}, \cite{sto92}, \cite{javi94}.

Only recently rigorous studies of the spectral properties of random
operators of the form \xref{2} have appeared \cite{mapu92},
\cite{mat93}, \cite{brhu93}, \cite{uek94}, \cite{wan94},
\cite{cohi94}, \cite{doma95a}, \cite{doma95b}, \cite{brhuPREP}, but
see also the related earlier work \cite{gus77}.  The self-averaging
property of the integrated density of states has been established
under fairly general conditions and various of its asymptotic
properties have been considered. Also the existence of localized
states has been proven \cite{cohi94}, \cite{doma95a}, \cite{doma95b}.

In this note we are concerned with the Schr\"odinger operator \xref{2}
in two dimensions with a Poisson random potential determined by a
non-negative single-site potential $\ssp$, decaying algebraically at
infinity. That is,
\be{1.2} 
   V_\omega(x) = \sum_j \ssp\!\left(x-p_\omega(j)\right), 
\ee 
where $p_\omega(j)$ are random points in the Euclidean plane $\rz^2$
distributed in accordance with Poisson's law.  We will show how
rigorous versions of results in \cite{brhe89} and \cite{brhe91} can be
used to find the leading asymptotic behaviour of the integrated
density of states as the energy approaches the infimum of the
spectrum from above.

In the zero-field case the asymptotic behaviour changes its character
depending on the decay of the single-site potential at infinity. For
slow decay it is governed solely by the potential energy \cite{pas77},
that is, by classical effects. For rapid decay the kinetic energy also
becomes relevant leading to genuine quantum effects. The form of the
latter asymptotic behaviour has been discovered by Lifshitz
\cite{lif64}, \cite{lif67}, \cite{ligr88}. Convincing arguments for
the validity of Lifshitz' conjecture were given by Friedberg and
Luttinger \cite{frlu75}, \cite{lut76}.  Its rigorous proof
\cite{dova75}, \cite{nak77}, \cite{pas77} relies on Donsker's and
Varadhan's involved large-deviation results \cite{dova75},
\cite[Section~4.3]{dest89} about the long-time asymptotics of certain
Wiener integrals.

For the case of non-zero constant magnetic field we will show that the
low-energy tail is always given by the classical result irrespective
of the decay of the single-site potential. Basically, this is due to
the fact that in this case the ground state of the unperturbed
Schr\"odinger operator $H(0)$ consists of square-integrable functions.

The distance between successive eigenvalues of $H(0)$, commonly
referred to as Landau levels, and the degeneracy per area of each
eigenvalue increase linearly with the strength of the magnetic field.
For a fixed concentration of non-interacting electrons and
sufficiently strong field it is therefore physically reasonable to
investigate only the restriction $E_0H(V_\omega)E_0$ to the eigenspace
$E_0\L^2(\rz^2)$ of the lowest Landau level instead of the full
Schr\"odinger operator $H(V_\omega)$. For a rigorous discussion of
this point see \cite{mapu92}, \cite{brhu93}, \cite{brhuPREP}.
Remarkably, Wegner \cite{weg83} succeeded in calculating the
corresponding restricted integrated density of states for the case of
a delta-correlated Gaussian random potential. A rigorous version of
this derivation is given in \cite{mapu92}. Wegner's result was quickly
extended to general delta-correlated random potentials by Br\'ezin,
Gross and Itzykson \cite{brgr84}, including Poisson potentials, the
single-site potential of which being a Dirac delta function. The
calculation in \cite{brgr84} relies on the resummation of a suitable
representation of an averaged Neumann series. A non-perturbative
derivation was given by Klein and Perez \cite{klpe85}.

We will show that the restriction to any Landau level does not alter
the leading asymptotic behaviour of the integrated density of states
at the lower spectral boundary for algebraically decaying single-site
potentials. This behaviour is in sharp contrast to that for
delta-correlated Poisson potentials, where it can happen that the
integrated density of states is not continuous at the infimum of the
spectrum.

Although we will only discuss the two-dimensional case, our result
generalizes to even dimensions and a non-degenerate magnetic-field
tensor, that is, to the case where the magnetic-field tensor
$(B_{jk})$ is constant and has full rank.
%-----------------------------------------------------------------------------
\abschnitt{II}{Statement of the Result}
%-----------------------------------------------------------------------------
We will assume throughout that the random potential $V_\omega$ is the
{\bf Poisson random potential} \cite[Example~1.15(d)]{pafi92} with
concentration $\rho > 0$ and {\bf non-negative single-site potential}
$\ssp:\rz^2\to[0,\infty]$, where $\ssp$ decays algebraically and 
integrably at infinity, that is
\be{2.1}
   \ssp\geq 0,\quad   
   \lim_{|x| \to \infty } |x|^\alpha \ssp(x)
   = \mu , \quad 0<\mu<\infty ,\; \alpha >2.
\ee
In addition, for technical reasons we cannot allow for too strong
local singularities. Therefore we will assume besides \xref{2.1} either 
\be{2.2a}
   \ssp\in \L^2_{\hbox{\scriptsize loc}}(\rz^2)
\ee
or, more restrictively,
\be{2.2b}
   \ssp\in \L^\infty_{\hbox{\scriptsize loc}}(\rz^2).
\ee
%--------------------------------------------------------------------------
\begin{structure}{2-1}{Remarks}\einruecken
\knoedel{iv}
   Either of the conditions ensure that $V_\omega$ is an ergodic measurable
   non-negative random potential, where ergodicity is meant with respect
   to the group of spatial translations in $\rz^2$. 
\knoedel{v}
   We note that \xref{2.1} together with \xref{2.2a} implies
   \be{2.2c}
      \| \ssp \|_p < \infty 
      \quad\hbox{ for all } 1\le p\le 2,
   \ee
   while \xref{2.1} together with \xref{2.2b} implies
   \be{2.2d}
      \| \ssp \|_p \leq 
      \left( \| \ssp \|_1 \right)^{1/p}
      \left( \| \ssp \|_\infty \right)^{(p-1)/p} < \infty
      \quad\hbox{ for all } 1\le p\le \infty .
   \ee
   Here as usual 
   \be{2.2e}
      \|f\|_p \defeq
      \left\{\;\matrix{\noalign{\vskip 5pt}\Ds
         \left(\;
            \int |f(y)|^p\,\d y
         \right)^{1/p}
         \hfill & &\Ds
         p<\infty
      \hfill\cr\Ds
         & \vbox to 0pt{\vss\hbox{for}} &
      \cr\Ds
         \esssup_{y\in\rz^2} |f(y)|
         \hfill & &\Ds
         p=\infty
      \hfill\cr\noalign{\vskip 5pt}}\right.
   \ee
   denotes the norm of $f\in\L^p(\rz^2)$.

   Let $\E$ denote the expectation with respect to the random potential.
   By \xref{2.2c} or \xref{2.2d} one has the finiteness
   \be{5a}
      \E\!\left[\left(V_\omega(x)\right)^2\right]
      =
      \varrho \int\left( \ssp(y)\right)^2\d y
      + 
      \left(\varrho \int \ssp(y)\,\d y\right)^2
      <\infty
   \ee
   of the potential's second moment at any point $x\in\rz^2$. 
   This implies that the potential $V_\omega$ is almost 
   surely locally square-integrable and, moreover, that the mapping
   \be{5b}
      x\mapsto
      \int_{|x-y|\le 1} \left( V_\omega(y) \right)^2 \d y
   \ee
   is almost surely polynomially bounded as $\left|x\right|\to\infty$, confer
   \cite[Proof of Theorem~5.1]{kir89}.

   By \xref{2.2d} one has the finiteness
   \be{2.4a}
      \matrix{\noalign{\vskip5.25pt}\Ds
         \E\!\left[\left(V_\omega(x)\right)^k\right]
      &\Ds\leq&\Ds
         k!\left(\|\ssp\|_\infty\right)^k
         \left(
            \E\!\left[\e^{V_\omega(x)/\|\ssp\|_\infty}\right] - 1
         \right)
      \hfill\cr\noalign{\vskip5.25pt}\Ds &\Ds =&\Ds
         k!\left(\|\ssp\|_\infty\right)^k
         \left(
            \exp\!\left\{
               \varrho\int\left(
                  \e^{\ssp(y)/\|\ssp\|_\infty}-1
               \right)\d y
            \right\} - 1
         \right)
      \hfill\cr\noalign{\vskip5.25pt}\Ds &\Ds\leq&\Ds
         k!\left(\|\ssp\|_\infty\right)^k
         \left(
            \exp\!\left\{
               \varrho \, \frac{\|\ssp\|_1}{\|\ssp\|_\infty} \, (\e-1)
            \right\} - 1
         \right)
      \hfill\cr\noalign{\vskip5.25pt}}
   \ee
   of the $k$-th moment for all positive integers $k\in\nz$.
\knoedel{ii}
   The Laplace characteristic functional of $V_\omega$
   \be{5}
       \E\!\left[\e^{-\int J(x) V_\omega(x)\, \d x}\right]=
       \exp\!\left\{
          -\rho \int \left(1-\e^{-\int J(x-y) \, \ssp(y)\,\d y}\right)\d x
       \right\} 
   \ee
   is well-defined for all non-negative $J \in {\cal S} (\rz^2)$,
   confer for example \cite[Proposition~1.16]{pafi92}. Here we have
   introduced ${\cal S}(\rz^2)$ for the Schwartz space of rapidly decreasing
   arbitrarily-often-differentiable complex-valued functions on $\rz^2$.
\knoedel{i}
   One can look upon $V_\omega$ as given by
   \be{5c}
      V_\omega(x) = \int \ssp(x-y)\,\d m_\omega(y) ,
   \ee
   where $m_\omega$ is the Poisson random measure on $\rz^2$ with  
   concentration $\varrho$. Since $m_\omega$ is purely atomic, 
   $V_\omega$ can be interpreted physically as the potential generated 
   by uniformly distributed non-interacting impurities, the influence of 
   each impurity being described by the same repulsive potential 
   $\ssp$, see \xref{1.2}.
\end{structure}
%--------------------------------------------------------------------------
In the sequel we suppose $B>0$ and the vector potential \be{7}
  A(x) \defeq
  \frac{B}{2} {x_2 \choose -x_1}, \quad x\eqdef{x_1 \choose x_2},
\ee
given in the symmetric gauge. Then the {\bf Schr\"odinger operator}
\xref{2} is almost surely essentially self-adjoint on ${\cal S}(\rz^2)$
provided \xref{2.1} and \xref{2.2a} hold.
This follows from \cite[Theorem~1.15]{cyfr87} and the fact that
$V_\omega\psi\in\L^2(\rz^2)$ for all $\psi\in{\cal S}(\rz^2)$ almost
surely because of Remark~\ref{2-1-v)}. On account of gauge equivalence
\cite{lei83}, with the choice \xref{7} there is no loss of generality for
the description of a constant magnetic field.

The spectral resolution of the unperturbed part dates back to Landau
\cite{lan30} and is given by
\be{8}
   H(0) = \sum^\infty _{n=0} \eps_n \, E_n ,\quad
   \eps_n  \defeq \left(n+ \frac{1}{2}\right)B ,
\ee
where the eigenvalues $\eps_n$ are called Landau levels and the
corresponding infinite-dimensional eigenprojectors $E_n$ are integral
operators with kernels given by
\be{9}
   E_n(x,y) \defeq
   \frac{B}{2\pi} \,
   \e^{\frac{iB}{2} (x_1y_2-x_2y_1)-\frac{B}{4}(x-y)^2} \,
   {\rm L}_n\!\left(\frac{B}{2}(x-y)^2\right).
\ee
Here ${\rm L}_n$ is the $n$-th Laguerre polynomial 
\cite[Section~8.97]{grry80}.

The explicit formula for the integral kernel of $E_n$ shows that
$E_n(x,\bullet)\in{\cal S}(\rz^2)$. Since ${\cal S}(\rz^2)$ is stable
under convolution, this implies that ${\cal S}(\rz^2)\cap
E_n\L^2(\rz^2)\subset E_n{\cal S}(\rz^2)\subset {\cal S}(\rz^2)$.
Furthermore, $E_n{\cal S}(\rz^2)$ is dense in $E_n\L^2(\rz^2)$,
because ${\cal S}(\rz^2)$ is dense in $\L^2(\rz^2)$.  Taking finally
into account, that the mapping \xref{5b} is almost surely polynomially
bounded, we conclude that the Schr\"odinger operator restricted to the
eigenspace of the $n$-th Landau level, in symbols
$E_nH(V_\omega)E_n=\eps_n E_n + E_n V_\omega E_n$, and all its
natural powers $\left(E_nH(V_\omega)E_n\right)^k$, $k\in\nz$, are
almost surely well-defined non-negative operators from ${\cal S}(\rz^2)$
to $\L^2(\rz^2)$.

As for the essential self-adjointness of $E_nH(V_\omega)E_n$ on 
${\cal S}(\rz^2)$, we were not able to prove it under the assumptions
\xref{2.1} and \xref{2.2a}. However, under the stronger assumptions
\xref{2.1} and \xref{2.2b} a proof can be taylored along the lines
given in the proof of Theorem~2.1 in \cite{doma95a}. By appealing to
Nelson's analytic-vector theorem \cite[Theorem~X.39]{resi75} and to the 
fact that ${\cal S}(\rz^2)$ contains a countable dense subset it suffices
to check that 
\be{2.9a}
   \sum_{k=0}^\infty \frac{t^k}{k!}
   \left( 
      \E\!\left[\left(\left\|
         \left(E_nV_\omega E_n\right)^k\psi
      \right\|_2\right)^2\right]
   \right)^{1/2}
   < \infty
\ee
for all $\psi\in{\cal S}(\rz^2)$ and some $t>0$. We start by observing 
that there is a constant $c_n(B)$ such that 
\be{2.9b}
   \left| E_n(x,y) \right| \leq
   c_n(B) \: \frac{B}{8\pi} \, 
   \e^{-\frac{B}{8} (x-y)^2}
\ee
for all $x,y\in\rz^2$. Moreover, by an iterated version of H\"older's 
inequality and translation invariance one has
\be{2.9c}
   \E\!\left[
      \prod_{j=1}^{2k} V_\omega(x^{(j)})
   \right]
   \leq
   \E\!\left[
      \left( V_\omega(0) \right)^{2k}
   \right]
\ee
for all $x^{(1)},\dots,x^{(2k)}\in\rz^2$.
Copying now the steps between Equations (2.15) and (2.20) in \cite{doma95a}
our estimates \xref{2.9b}, \xref{2.9c} and \xref{2.4a} yield for all 
$k\in\nz$ 
\be{2.9d}
   \matrix{\noalign{\vskip5.25pt}\Ds
   \E\!\left[\left(\left\|
      \left(E_nV_\omega E_n\right)^k\psi
   \right\|_2\right)^2\right]
   &\Ds \!\!\leq\!\! & \Ds 
   (2k)!\,\left(c_n(B) \, \|\ssp\|_\infty\right)^{2k}\,
   \hfill\cr\noalign{\vskip5.25pt}\Ds &\Ds &\Ds \times
   \frac{B c_n(B) \left(\|\psi\|_1\right)^2}{8\pi(2k+1)}
   \left(
      \exp\!\left\{
         \varrho \, \frac{\|\ssp\|_1}{\|\ssp\|_\infty} \, (\e-1)
      \right\} -1
   \right).
   \hfill\cr\noalign{\vskip5.25pt}}
\ee
As a consequence, \xref{2.9a} is valid for all 
$0\leq t < \left(2 \, c_n(B) \, \|\ssp\|_\infty\right)^{-1}$.

In the following we are allowed and will understand the unrestricted
operators $H(V_\omega)$ and the restricted operators $E_n H(V_\omega)
E_n$ as being almost surely essentially self-adjoint on ${\cal
  S}(\rz^2)$ under the imposed assumptions \xref{2.1}, \xref{2.2a} and
\xref{2.1}, \xref{2.2b}, repectively.

Let $\rz\ni\lambda\mapsto P_\lambda(X) = \Theta(\lambda-X)$ denote the
projection-valued measure for the self-adjoint operator $X$. Here
$\Theta$ is Heaviside's unit-step function: $\Theta(a)=0$ for $a<0$
and $\Theta(a)=1$ for $a\geq 0$. Our objects of study, the 
{\bf integrated density of states} $N$ and the {\bf {\boldmath $n$}-th 
restricted integrated density of states} $R_n$, are defined by
\be{10}
   N(\lambda) \defeq \E[P_\lambda(H(V_\omega))(x,x)]
\ee
and
\be{11}
   R_n(\lambda) \defeq
   \E\!\left[
       (E_n\,P_\lambda(E_nH(V_\omega)E_n)\,E_n)(x,x)
   \right],
\ee
respectively.
%--------------------------------------------------------------------------
\begin{structure}{2-2}{Remarks}\einruecken
\knoedel{i}
   The integral kernel $(x,y)\mapsto P_\lambda(H(V_\omega))(x,y)$ of the
   spectral projection $P_\lambda(H(V_\omega))$ almost surely exists and
   is jointly continuous in $(x,y)$. This can be seen from
   \cite[Section~6]{brhu94} using Remark~\ref{2-1-v)}. For vanishing
   vector potential this result is standard \cite[Theorem~B.7.1]{sim82}.
   The above continuity assures that \xref{10} is well-defined. Since
   $V_\omega$ is translation invariant, the right-hand side of \xref{10} is
   independent of $x \in \rz^2$.
\knoedel{ii}
   In \cite[Proposition~3.2]{uek94}, \cite{brhuPREP} it is shown that the 
   more familiar and physically
   reasonable way of defining the integrated density of states by means
   of a macroscopic limit yields \xref{10}. This amounts to first
   restricting the Schr\"odinger operator $H(V_\omega)$  to a finite
   region -- a square with Dirichlet boundary conditions, say. Then one
   defines the integrated density of states of the finite system to be
   the number of eigenvalues below $\lambda$ divided by the region's
   area. Finally, one proves that this quantity becomes non-random in the
   macroscopic limit, which is usually summarized as the self-averaging
   property of the integrated density of states.
\knoedel{iii}
   The Cauchy-Schwarz inequality, $E_n^2=E_n$ and $E_n(x,x)=B/(2\pi)$  
   yield for all $\psi\in\L^2(\rz^2)$
   \be{13}
       \|E_n\psi\|_\infty \le
       \left(\frac{B}{2\pi} \left<\psi,\psi\right>\right)^{1/2}
       =
       \left(\frac{B}{2\pi}\right)^{1/2}\, \|\psi\|_2,
   \ee
   where, as usual, $\left<\bullet,\bullet\right>$ symbolizes
   the standard scalar product for $\L^2(\rz^2)$. Therefore, the
   restriction $E_nXE_n$ of any bounded self-adjoint operator $X$  
   on $\L^2(\rz^2)$ to the eigenspace of the
   $n$-th Landau level is a Carleman integral operator
   \cite[Corollary~A.1.2]{sim82}. Since $E_n(x,y)$ is smoothing and a
   projection, the integral kernel $(x,y)\mapsto(E_nXE_n)(x,y)$ is
   jointly continuous in $(x,y)$. Hence \xref{11} is well-defined, too.
   Furthermore, the right-hand side of \xref{11} is independent of
   $x\in\rz^2$ due to the translation invariance of $V_\omega$, see
   the Appendix for details.
\knoedel{iv}
   On physical grounds $R_0$ should be a reasonable approximation to $N$
   for strong magnetic fields. This is given a precise meaning in
   \cite[Proposition~1]{mapu92}, \cite[Theorem~5]{brhu93} and \cite{brhuPREP}.
   For the significance of $R_n$ for general $n$ see \cite{brhe91}, 
   \cite{brhu93}.
\knoedel{v} 
   In the reasoning in the above Remarks {\it i)}\/ and {\it iii)}\/ we
   have swept measurability questions with respect to $\omega$ under the
   rug as we will do in the sequel. These problems, however, can be fixed
   by the methods of \cite{kima82} or \cite[Sections~V.1, V.3]{cala90}.
\end{structure}
%--------------------------------------------------------------------
Our aim is to identify the asymptotics of $N(\lambda)$ and
$R_n(\lambda)$ as $\lambda$ approaches the lower spectral boundary,
that is, $\lambda\downarrow \hbox{\rm inf spec} (H(V_\omega))$ or
$\lambda\downarrow \hbox{\rm inf spec} (E_nH(V_\omega)E_n)$,
respectively.  By employing the so-called magnetic translations
\cite{zak64}, \cite{klpe85}, see Equation \xref{A.18} below, standard
arguments \cite{kima82}, \cite{kir89} show that both $H(V_\omega)$ and
$E_nH(V_\omega)E_n$ are ergodic families of operators \cite{uek94},
\cite{wan94}, \cite{cohi94}, \cite{doma95a}, \cite{brhuPREP}.
Therefore, their spectra are non-random quantities, see
\cite[Theorem~2.1]{uek94}, \cite[Theorem~1]{kima82} or
\cite[Theorem~4.3.1]{kir89}.  It will turn out that $\hbox{\rm inf
  spec} (H(V_\omega)) = \eps_0$ and $\hbox{\rm inf spec}
(E_nH(V_\omega)E_n) = \eps_n$ as expected. Our result is the
following.
%--------------------------------------------------------------------
\begin{structure}{2-3}{Theorem}\einruecken\it
   Under the assumptions \xref{2.1} and \xref{2.2a} one has for all $B>0$
   \be{15}
      \lim_{\lambda \downarrow 0}
      \lambda^{2/(\alpha -2)} \ln N(\eps_0 + \lambda)
      = - C(\alpha,\mu,\varrho)
   \ee
   for the integrated density of states $N$ and, similarly,
   under the assumptions \xref{2.1} and \xref{2.2b} 
   \be{15a}
      \lim_{\lambda \downarrow 0}
      \lambda^{2/(\alpha -2)} \ln R_n(\eps_n + \lambda)
      = - C(\alpha,\mu,\varrho)
   \ee
   for the $n$-th restricted integrated density of states $R_n$,
   $n\in\nz\cup\{0\}$.\\
   Here we have set
   \be{16}
      C(\alpha,\mu,\varrho) \defeq
      \frac{1}{2} (\alpha -2)\,
      \mu^{2/(\alpha -2)}
      \left(
         \frac{2\pi\varrho}{\alpha} \,
         \Gamma\!\left(\frac{\alpha-2}{\alpha}\right)
      \right)^{\alpha/(\alpha -2)}
      > 0,
   \ee
   where $\Gamma$ denotes Euler's gamma function.
\end{structure}
%--------------------------------------------------------------------------
\begin{structure}{2-4}{Remarks}\einruecken
\knoedel{i}
   The result for the unrestricted integrated density of states $N$
   should be compared with the case $B=0$. The asymptotic decay at the
   lower spectral boundary coincides with the behaviour for $B=0$, if
   $(d=)\,2<\alpha<4\,(=d+2)$, that is, for slowly decaying single-site
   potential, see \cite{pas77} or \cite[Corollary~9.14]{pafi92}, but it
   differs in the case $\alpha>4\,(=d+2)$, see \cite{nak77},
   \cite{pas77} or \cite[Theorem~10.2]{pafi92}. 

   This is plausible from the Rayleigh-Ritz-like variational principle
   due to Luttinger \cite{lut76} and Pastur \cite{pas77}, see also 
   Equation (17.3) and Chapter~21 in \cite{ligr88}. For $B=0$ and 
   $\alpha>4$ the optimal wavefunction becomes too sharply localized 
   so that the unperturbed (kinetic) energy begins to play a significant 
   r\^ole. For $B>0$ the contribution of the unperturbed energy relative 
   to the ground-state energy $\eps_0$ can always be kept zero by
   choosing one of the square-integrable ground-state wavefunctions of 
   $H(0)$. 
\knoedel{ii}
   The lowest restricted integrated density of states $R_0$ for a
   delta-correlated Poisson potential, corresponding to
   $\ssp(x)=\nu\delta(x)$, $\nu>0$, is known exactly
   \cite{brgr84}, \cite{klpe85}, not only in the low-energy tail. 
   If the mean number $2\pi\varrho/B$ of impurities over
   the spatial extent $\pi l^2$ of the ground-state wavefunctions 
   of $H(0)$,
   \be{16a}
      l^2\defeq \frac{1}{E_0(y,y)}
      \int E_0(y,x)\,(x-y)^2\,E_0(x,y)\, \d x = \frac{2}{B} ,
   \ee
   is smaller than one, $R_0$ exhibits a jump at $\eps_0$, the height
   of which being proportional to $1-2\pi\varrho/B$. This is
   plausible, because in the case $2\pi\varrho/B<1$ one can imagine
   that, effectively, only a fraction of the ground-state
   wavefunctions is affected by the impurities \cite{brgr84}. 
   This reasoning does not seem applicable to algebraically
   decaying impurity potentials, in accordance with our result that
   $R_n$ is continuous at $\eps_n$.
\knoedel{iii}
   For the reader's convenience, we present the generalizations of 
   \xref{15} and \xref{15a} to even space dimensions $d\geq 2$ 
   \be{16b}
      \matrix{\noalign{\vskip\displaystylematrixsep}\Ds &\Ds
         \lim_{\lambda\downarrow 0}
         \lambda^{d/(\alpha-d)} \ln N(\eps_0+\lambda)
         =
         \lim_{\lambda\downarrow 0}
         \lambda^{d/(\alpha-d)} \ln R_n(\eps_n+\lambda)
      \hfill\widecr\Ds\hfill = &\Ds
         -\,\frac{\alpha-d}{d}\,
         \mu^{d/(\alpha-d)}
         \left(
            \frac{\varrho}{\alpha} \,
            \frac{2\pi^{d/2}}{\Gamma(d/2)} \,
            \Gamma\!\left(\frac{\alpha-d}{\alpha}\right)
         \right)^{\alpha/(\alpha-d)}.
      \hfill\widecr}
   \ee
   Here it is assumed that $\alpha>d$ and that the magnetic-field 
   tensor $(B_{jk})$ is constant and has full rank.
\knoedel{iv}
   It would be interesting to compute subleading corrections to
   \xref{16b} as Luttinger and Waxler \cite{luwa87} did for zero
   magnetic field and, of course, $d<\alpha<d+2$. In contrast to the
   leading term given by \xref{16b} we expect these corrections to
   depend on the field. 
\end{structure}
%-----------------------------------------------------------------------------
\abschnitt{III}{Proof}
%-----------------------------------------------------------------------------
For the proof of Theorem \ref{2-3} we follow the strategy in
\cite{pas77}.  Instead of $N(\lambda)$ we investigate its Laplace
transform and use a Tauberian argument \cite[Theorem~9.7]{pafi92}.
Since $H(V_\omega)$ is bounded below by $\eps_0$, this argument shows
that \xref{15} is equivalent to

\be{17}
   \lim_{t\to\infty}
      t^{-2/\alpha} \ln \widetilde N (t)
      =
      - \pi \varrho \mu^{2/\alpha}
        \Gamma\!\left(\frac{\alpha-2}{\alpha}\right)
      =
      - \varrho \mu^{2/\alpha}
        \int \left( 1-\e^{- |x|^{-\alpha}}\right)\, \d x
      ,
\ee
where
\be{18}
    \widetilde N (t) :=
    \int \e^{-t\lambda} \, \d N(\eps_0+\lambda) =
    \e^{t\eps_0} \int \e^{-t\lambda} \, \d N(\lambda)
    , \quad t>0,
\ee
is the shifted Laplace transform of $N$. Analogously we define
\be{18a}
    \widetilde R_n (t) :=
    \int \e^{-t\lambda} \, \d R_n(\eps_n+\lambda) =
    \e^{t\eps_n} \int \e^{-t\lambda} \, \d R_n(\lambda)
    , \quad t>0,
\ee
to be the shifted Laplace transform of $R_n$. Then \xref{15a} is
equivalent to \xref{17} with $\widetilde R_n$ replacing $\widetilde N$.

To establish the leading asymptotic behaviour of $\widetilde N(t)$ and
$\widetilde R_n(t)$ for large $t$ we use the following
%--------------------------------------------------------------------
\begin{structure}{3-1}{Basic Inequalities}\einruecken\it
Let
\be{23}
   \phi_n(\bullet) \defeq 
   \left(\frac{2\pi}{B}\right)^{1/2}\! E_n(\bullet,0).
\ee
Then
\be{19}
   \frac{1 }{2 \pi t}
   \E\!\left[ \e^{-t\left<\phi_0, V_\omega \phi_0\right>} \right]
   \le
   \widetilde N(t),
\ee
\be{20}
   \widetilde N(t)
   \le
   \frac{B\,\e^{t\eps_0}}{4\pi \sinh(t\eps_0)}
   \E\!\left[ \e^{-t V_\omega(0)} \right],
\ee
and
\be{21}
   \frac{B}{2 \pi}
   \E\!\left[ \e^{-t\left<\phi_n, V_\omega \phi_n\right>} \right]
   \le
   \widetilde R_n(t),
\ee
\be{22}
   \widetilde R_n(t) \le
   \frac{B}{2\pi} \E\!\left[ \e^{-t V_\omega(0)} \right].
\ee%
\end{structure}%
%--------------------------------------------------------------------
\unskip%
%--------------------------------------------------------------------
\begin{structure}{3-2}{Remarks}\einruecken%
\knoedel{i}%
   One can infer from \xref{9} that $\|\phi_n\|_2=1$ and
   $\phi_n\in{\cal S}(\rz^2)$. Therefore, the left-hand sides of
   \xref{19} and \xref{21} are well-defined, because almost surely
   $V_\omega\in \L^2_{\hbox{\scriptsize loc}}(\rz^2)$ and the 
   mapping \xref{5b} is polynomially bounded.
\knoedel{ii}
   The bound \xref{20} is a generalization of the classical upper bound
   for $\widetilde{N}$ \cite{pas77}, \cite[Theorem~9.6]{pafi92} to
   non-zero magnetic fields. The essential ingredient for its derivation
   is the Golden-Thompson inequality. The bounds
   \xref{21} as well as \xref{22} stem from the Jensen-Peierls
   inequality. The inequality \xref{19} is a
   variant of an inequality due to Berezin, Lieb and Luttinger, which
   in turn follows from the Jensen-Peierls inequality.
   In \cite{pas77} and
   \cite[Theorem~9.5]{pafi92} a lower bound similar to \xref{19} is
   proven. It has the advantage of holding for more general families of
   random operators but allows for functions $\phi_0$ with compact
   support only. For our proof it is essential, however, to allow for
   $\phi_0$ as given in \xref{23} with unbounded support. 
\knoedel{iii}
   Non-rigorous derivations of the above bounds can be found in
   \cite{brhe89} and \cite{brhe91}.
\knoedel{iiii}
   A rigorous derivation of \xref{20} for the case of a Gaussian random 
   potential with a Gaussian covariance function can be read off from 
   Equations (2.3), (2.8) and (2.10) in \cite{mapu92}.
\knoedel{iv}
   The above bounds are to some extent special cases of those presented
   in \cite{brhuPREP} for more general than Poisson random potentials. Our
   lines of reasoning are closely related to those in \cite{brhuPREP} but
   are considerably simplified by the fact that $V_\omega\ge 0$ almost
   surely. We have banished the proofs of the Basic Inequalities into
   the Appendix because on the one hand they follow the plan of
   \cite{brhe89} and \cite{brhe91} and on the other hand there are some
   unwieldy technicalities involved in supplying the missing rigour.
\end{structure}
%--------------------------------------------------------------------
The Basic Inequalities provide us with asymptotically coinciding upper
and lower bounds for the shifted Laplace transforms $\widetilde N$ and
$\widetilde R_n$.
%--------------------------------------------------------------------
\subsubsection*{Upper bound}
The inequality \xref{22} and Remark \ref{2-1-ii)} imply
\be{24}
  \widetilde R_n(t)
  \le
  \frac{B}{2\pi}
  \exp\!\left\{
     -\varrho \int\left( 1-\e^{-t\ssp(x)}\right)\d x
  \right\} .
\ee
Therefore we have
\be{25}
   \limsup_{t \to \infty}t^{-2/\alpha} \ln \widetilde R_n(t)
   \le -\varrho \liminf_{t \to \infty}t^{-2/\alpha}
   \int \left(
      1-\e^{-t \ssp(x)}
   \right)\d x .
\ee
On the right-hand side of \xref{25} we substitute $x\mapsto
(\mu t)^{1/\alpha}x$ and use Fatou's lemma to interchange the limit and
the integration. Since
\be{26}
   \lim_{t\to\infty}t\,\ssp\!\left((\mu t)^{1/\alpha}x\right)=|x|^{-\alpha}
   ,\quad x\not=0,
\ee
by \xref{2.1}, we conclude
\be{27}
   \limsup_{t \to \infty}t^{-2/\alpha} \ln \widetilde R_n(t)
   \le
   - \varrho \mu^{2/\alpha}
   \int \left(
      1-\e^{- |x|^{-\alpha}}
   \right) \d x .
\ee
Note that this is also true with $\widetilde N$ replacing $\widetilde
R_n$, because the different prefactors in the upper bounds \xref{20}
and \xref{22} coincide asymptotically.\halmos
%--------------------------------------------------------------------
\subsubsection*{Lower bound}
By means of Remark \ref{2-1-ii)} the inequality \xref{21} reads more 
explicitly
\be{28}
   \frac{B}{2 \pi}
   \exp\left\{
      - \rho \int \left(
         1-\e^{
             -t\int |\phi_n(x-y)|^2 \ssp(y) \, \d y
         }
      \right) \, \d x
   \right\}
  \le 
  \widetilde R_n(t) .
\ee
With the help of the substitution $x\mapsto(\mu t)^{1/\alpha}x$ this
yields 
\be{29}
    - \varrho \mu^{2/\alpha}
    \limsup_{t \to \infty} \int \left(
       1-
       \e^{- \int
          \delta_t(x-y) \,t\,\ssp\left((\mu t)^{1/\alpha}y\right)
       \, \d y }
    \right) \d x 
  \le
    \liminf_{t \to \infty} t^{-2/\alpha}
    \ln \widetilde R_n(t)
  ,
\ee
where we have introduced the one-parameter family 
$\{\delta_t\}_{t>0}\subset{\cal S}(\rz^2)$
of probability densities on $\rz^2$ 
\be{30}
   x\mapsto\delta_t(x)
   \defeq
   \frac{B\,(\mu t)^{2/\alpha}}{2\pi}\,
   \e^{-B\,(\mu t)^{2/\alpha}x^2/2}
   \left[ {\rm L}_n\!\left(B\,(\mu t)^{2/\alpha} x^2/2\right)
   \right]^2.
\ee
The Fourier representation
\be{30a}
   \delta_t(x) =
   \frac{1}{(2\pi)^2}\int \e^{ikx} \,
   \e^{-k^2/\left(2B\,(\mu t)^{2/\alpha}\right)}
   \left[ {\rm L}_n\!\left(k^2/(2B\,(\mu t)^{2/\alpha})\right)
   \right]^2
   \d k,
\ee
which may be verified with the help of \cite[Equation~7.377]{grry80},
shows that $\delta_t$ approximates Dirac's delta function as
$t\to\infty$.  Therefore one can check
\be{31}
    \limsup_{t \to \infty}
    \int \delta_t(x-y) \,t\,\ssp\!\left((\mu t)^{1/\alpha}y\right)
       \, \d y 
    \le
    |x|^{-\alpha} ,\quad x \neq 0,
\ee
using \xref{26} and the fact that $\ssp$ is both integrable and
square-integrable. According to Fatou's lemma this suffices for
the validity of
\be{32}
   - \varrho \mu^{2/\alpha}
   \int \left(
      1-\e^{- |x|^{-\alpha}}
   \right) \d x 
   \le
   \liminf_{t \to \infty}t^{-2/\alpha} \ln \widetilde R_n(t)
   .
\ee
To obtain the same estimate with $\widetilde N$ replacing $\widetilde
R_n$ one only has to specialize to $n=0$ and to note that the
differing prefactors in \xref{19} and \xref{21} become both irrelevant
on the logarithmic scale.\halmos
%-----------------------------------------------------------------------------
\begin{appendix}
%-----------------------------------------------------------------------------
\abschnitt{A}{Proofs of the Basic Inequalities}
\def\thesection{\Alph{section}}
%-----------------------------------------------------------------------------
The proofs of the bounds \xref{19} and \xref{20} for $\widetilde N$
rely on an approximation formula, which will be presented first.
%--------------------------------------------------------------------
\begin{structure}{A-1}{Approximation}\einruecken\it
Define for $\Omega\geq 0$
\be{A.1}
   \gdef\vreg{\hat V_{\omega,\Omega}}
   \vreg(x) \defeq V_\omega(x) + \frac{\Omega^2}{2}\, x^2.
\ee
Then for $t,\Omega>0$ the Euclidean propagator $\e^{-tH(\vreg)}$ is
almost surely of trace class and one has
\be{A.2}
   \widetilde N(t) =
   \lim_{\Omega\downarrow 0} \frac{\Omega^2t}{2\pi}\, \e^{t\eps_0}\,
   \E\!\left[ \tr\!\left\{ \e^{-tH(\vreg)} \right\}\right].
\ee
\end{structure}
%--------------------------------------------------------------------
\subsubsection*{Proof}
According to the Feynman-Kac-It\^o formula, see for example
\cite[Theorem 15.5]{sim79a}, $\vreg\in\L^2_{\hbox{\scriptsize
loc}}(\rz^2)$ and $\vreg\geq 0$ ensure that $\e^{-tH(\vreg)}$ has the
integral kernel
\be{A.3}
   \e^{-tH(\vreg)}(x,y) \defeq \frac{1}{2\pi t}\,
   \e^{-(x-y)^2/(2t)}\, 
   \int \e^{-S_t(\vreg|b)} \, \d\mu_{0,x}^{t,y}(b),
\ee
where $\mu_{0,x}^{t,y}$ denotes the probability measure associated with the
two-dimensional Brownian bridge from $b(0)=x$ to $b(t)=y$.
Here the potentials' part of the Euclidean action is given by
\be{A.5}
   S_t(\vreg|b) \defeq
   i\int_0^t A(b(s)) \, \d b(s) + \int_0^t \vreg(b(s)) \, \d s.
\ee
Since the Brownian bridge is a continuous semimartingale \cite[Example
V.6.3]{pro92}, the It\^o stochastic line integral in \xref{A.5} is 
well-defined \cite[Section~II.4]{pro92}.

In \cite[Section 6]{brhu94} it is proven that the right-hand side of
\xref{A.3} is jointly continuous in $(t,x,y)$, $t>0$, $x,y\in\rz^2$, and
we will get as a by-product in the proof of \xref{20} below that
$\e^{-tH(\vreg)}$ is of trace class. Thus \cite[Example~X.1.18]{kat84}
we may write
\be{A.6}
   \tr\!\left\{ \e^{-tH(\vreg)} \right\} = \frac{1}{2\pi t}\,
   \int \e^{-S_t(\vreg|b+x)} \, \d\mu_{0,0}^{t,0}(b)\, \d x,
\ee
where we have performed the rigid shift $b\mapsto b+x$.
Due to the translation invariance of $V_\omega$ we get with the
help of It\^o's formula \cite[Corollary to Theorem~II.32]{pro92}
\be{A.7}
   \E\!\left[ \tr\!\left\{ \e^{-tH(\vreg)} \right\}\right] =
   \frac{1}{\Omega^2t^2} \,
   \E\!\left[ \int \e^{-S_t(V_\omega|b)-\Omega^2t\,\sigma_t^2(b)/2} \,
              \d\mu_{0,0}^{t,0}(b)
   \right],
\ee
where
\be{A.8}
   \sigma_t^2(b) \defeq
   \int_0^t \left(b(s)\right)^2 \frac{\d s}{t} -
   \left(\int_0^t b(s) \frac{\d s}{t} \right)^2 \geq 0.
\ee
Since by $V_\omega\geq  0$ one has
\be{A.9}
   \left| \e^{-S_t(V_\omega|b)-\Omega^2t\,\sigma_t^2(b)/2} \right| \leq 1,
\ee
the theorem of dominated convergence is applicable and yields
\be{A.10}
   \lim_{\Omega\downarrow 0} \frac{\Omega^2t}{2\pi}
   \E\!\left[ \tr\!\left\{ \e^{-tH(\vreg)} \right\}\right] =
   \frac{1}{2\pi t}\,
   \E\!\left[ \int \e^{-S_t(V_\omega|b)}\,\d\mu_{0,0}^{t,0}(b) \right].
\ee
Employing again the Feynman-Kac-It\^o formula, for $\Omega=0$, the continuity of the
involved integral kernels and the translation invariance of the random
potential, one achieves by \xref{10} and \xref{18}
\be{A.11}
   \frac{1}{2\pi t}\,
   \E\!\left[ \int \e^{-S_t(V_\omega|b)}\,\d\mu_{0,0}^{t,0}(b) \right]
   = \e^{-t\eps_0} \, \widetilde N(t),
\ee
which concludes the proof of \xref{A.2}.\halmos
%--------------------------------------------------------------------
\subsubsection*{Proof of \xref{19}}
Let
\be{A.12}
   \psi_{p,q}(x) \defeq \e^{ip\left(x+\frac{q}{2}\right)} \,
   \phi_0(x-q).
\ee
Then $\left\{\psi_{p,q}\right\}_{p,q\in\rz^2}$ is nothing but the
standard overcomplete family of coherent states associated with the
Heisenberg-Weyl group generated from the ground state of a
two-dimensional isotropic harmonic oscillator. Since
$\e^{-tH(\vreg)}$ is almost surely of trace class whenever
$t,\Omega>0$, we may write, see \cite[Equation~1.13]{ber71} or
\cite[Theorem~A.1.2]{sim80}
\be{A.13}
   \frac{1}{(2\pi)^2} \int
   \left< \psi_{p,q}, \e^{-tH(\vreg)} \psi_{p,q} \right>
   \d p \, \d q
   = \tr\!\left\{ \e^{-tH(\vreg)} \right\}. 
\ee
Since $H(\vreg)$ is almost surely essentially self-adjoint on ${\cal
S}(\rz^2)$ and $\psi_{p,q}\in{\cal S}(\rz^2)$, the Jensen-Peierls
inequality \cite{ber72}, \cite[Section~8(c)]{sim79b} implies
\be{A.14}
   \e^{-t\left< \psi_{p,q},H(\vreg)\psi_{p,q} \right>}  \le
   \left< \psi_{p,q}, \e^{-tH(\vreg)} \psi_{p,q} \right>.
\ee
Due to the translation invariance of the random potential, we get by
inserting \xref{A.14} into \xref{A.13} after some calculation
\be{A.15}
   \frac{2\pi}{\Omega^2t} \, \e^{-\Omega^2t/B} \,
   \frac{\e^{-t\eps_0}}{2\pi t} \,
   \E\!\left[ \e^{-t\left<\phi_0,V_\omega\phi_0\right>} \right]
   \leq \E\!\left[ \tr\!\left\{ \e^{-tH(\vreg)} \right\}\right].
\ee
With the help of \xref{A.2} this proves \xref{19}.\halmos
%--------------------------------------------------------------------
\subsubsection*{Proof of \xref{20}}
First note that for $\Omega>0$ the operators $H(0)$, $\vreg$ and $H(\vreg)$
are almost surely essentially self-adjoint on ${\cal S}(\rz^2)$
and non-negative. Furthermore, $\e^{-tH(0)/2}\,\e^{-t\vreg/2}$ is
almost surely Hilbert-Schmidt, since \cite[Theorem~VI.23]{resi80}
\be{A.16}
   \int \left| \,\e^{-tH(0)/2}(x,y)\,\e^{-t\vreg(y)/2}\,
   \right|^2 \d x\,\d y
   =
   \frac{B}{4\pi\sinh(t\eps_0)} \, \int \e^{-t\vreg(y)}\, \d y
   < \infty .
\ee
Therefore, we can use the Golden-Thompson inequality in the version
given in the Corollary to \cite[Theorem XIII.103]{resi78} to conclude
that $\e^{-tH(\vreg)}$ is almost surely of trace class and its trace is
bounded from above by the squared Hilbert-Schmidt norm \xref{A.16}.
Taking the expectation with respect to the random potential, this shows
\be{A.17}
   \E\!\left[ \tr\!\left\{ \e^{-tH(\vreg)} \right\}\right] \leq
   \frac{2\pi}{\Omega^2 t} \,
   \frac{B}{4\pi\sinh(t\eps_0)} \, \E\!\left[ \e^{-t V_\omega(0)} \right].
\ee
Using \xref{A.2} once again, \xref{20} is proven.\halmos

%--------------------------------------------------------------------
As a preliminary to the proofs of \xref{21} and \xref{22} we show that
the $n$-th restricted integrated density of states $R_n$ is  independent
of $x\in\rz^2$. This is achieved with the help of the unitary
magnetic-translation operator $W_x$ defined by
\be{A.18}
   \left( W_x \, \psi \right) (y) \defeq
   \e^{\frac{iB}{2}\left(y_1x_2-y_2x_1\right)}\, \psi(y-x),
   \quad \psi \in\L^2(\rz^2),
\ee
because on the one hand the eigenprojectors $E_n$ are left invariant
under its action, that is
\be{A.19}
   W^\dagger_x \, E_n \, W_x = E_n,
\ee
and on the other hand the potential is simply shifted
\be{A.20}
   W^\dagger_x \, V_\omega \, W_x = V_\omega \circ T_x ,\quad
   T_xy\defeq y+x.
\ee
Using these properties together with the translation invariance of the
random potential, the right-hand side of \xref{11} is seen to be
independent of $x\in\rz^2$. For a complete discussion of the
magnetic-translation group associated with $H(0)$ see 
\cite{zak64}, \cite{klpe85}.
%--------------------------------------------------------------------
\subsubsection*{Proof of \xref{21}}
The combination of the definition \xref{11} for $x=0$ with \xref{18a}
and \xref{23} gives a more explicit expression
\be{A.21}
   \widetilde R_n(t) = \frac{B}{2\pi}
   \E\!\left[ \left<\phi_n,\e^{-tE_nV_\omega E_n}\, \phi_n\right>\right]
\ee
for the shifted Laplace transform of the $n$-th restricted integrated
density of states. Since $\phi_n\in{\cal S}(\rz^2)$ and $E_nV_\omega
E_n$ is almost surely essentially self-adjoint on ${\cal S}(\rz^2)$, we
may use the Jensen-Peierls inequality \cite{ber72},
\cite[Section~8(c)]{sim79b} and the fact that $E_n\phi_n=\phi_n$ to
estimate
\be{A.22}
\e^{-t\left<\phi_n,V_\omega\phi_n\right>} \leq
\left<\phi_n,\e^{-tE_nV_\omega E_n}\, \phi_n\right> .
\ee
The insertion of \xref{A.22} into \xref{A.21} completes the proof of
\xref{21}.\halmos

For the proof of \xref{22} we will use again an approximation formula
which we single out as
%--------------------------------------------------------------------
\begin{structure}{A-2}{Approximation}\einruecken\it
Define the centered Poisson potential, truncated outside a disk 
of radius $r>0$ about the origin,
\be{A.23}
   \gdef\Vreg{\check V_{\omega,r}}
   \Vreg(x) \defeq
   \Theta\!\left(r-\left|x\right|\right)
   \left( V_\omega(x) - \E\!\left[V_\omega(0)\right] \right).
\ee
Then
\be{A.24}
   \lim_{r\to\infty} \frac{1}{\pi r^2} \int_{\left|x\right|\leq r}
      \E\!\left[\left( E_n\,\e^{-tE_n\Vreg E_n}\,E_n
      \right)(x,x)\right]
   \d x =
   \e^{t\,\E\left[V_\omega(0)\right]}\, \widetilde R_n(t).
\ee
\end{structure}
%--------------------------------------------------------------------
\subsubsection*{Proof}
First we use for the magnetic-translation operator \xref{A.18} the properties
\xref{A.19} and \xref{A.20} with $\Vreg$ replacing $V_\omega$, the
definition \xref{23} of $\phi_n$ and the translation invariance of
$V_\omega$ to rewrite
\be{A.25}
   \matrix{\noalign{\vskip\displaystylematrixsep}\Ds &\Ds
      \gdef\Vregreg{\bar V_{\!\omega,r,\zeta}}
      \frac{1}{\pi r^2} \int_{\left|x\right|\leq r}
         \E\!\left[\left( E_n\,\e^{-tE_n\Vreg E_n}\,E_n
         \right)(x,x)\right]
      \d x
   \hfill\widecr\Ds\hfill = &\Ds
      \frac{B}{2\pi^2} \int_{\left|\zeta\right|\leq 1}
         \E\!\left[\left< \phi_n,\e^{-tE_n\Vregreg E_n}\phi_n
         \right>\right]
      \d\zeta .
   \hfill\widecr}
\ee
Here we have introduced
\be{A.26}
   \Vregreg(x) \defeq
   \Theta\!\left(1-\left|\zeta-\frac{x}{r}\right|\right)
   \left( V_\omega(x) - \E\!\left[V_\omega(0)\right] \right) .
\ee

As a next step we claim that for all $\left|\zeta\right|<1$ and almost
all $\omega$
\be{A.27}
   E_n \Vregreg E_n \stackrel{r\to\infty}{\longrightarrow}
   E_n \left( V_\omega - \E\!\left[V_\omega(0)\right] \right) E_n
\ee
in the strong resolvent sense. Since $E_n \Vregreg E_n$ is almost
surely essentially self-adjoint on ${\cal S}(\rz^2)$ for all $r>0$ and
so is the right-hand side of \xref{A.27}, it is sufficient
\cite[Theorem~VIII.25]{resi80} to check that for all $\psi\in{\cal
S}(\rz^2)$ and almost all $\omega$
\be{A.28}
   \left\| E_n \left( 
      \Vregreg - V_\omega + \E\!\left[V_\omega(0)\right] 
   \right) E_n\psi \right\|_2
   \stackrel{r\to\infty}{\longrightarrow} 0 .
\ee
Since $E_n$ is bounded and maps ${\cal S}(\rz^2)$
into itself, this follows from
\be{A.29}
   \matrix{\noalign{\vskip\displaystylematrixsep}\Ds &\Ds
      \int\left|
         \left( \Vregreg(x) - V_\omega(x) + \E\!\left[V_\omega(0)\right]
         \right) \psi(x)
      \right|^2 \d x
   \hfill\widecr\Ds\hfill \le &\Ds
      \int_{\left|x\right|>r\left(1-\left|\zeta\right|\right)}\left|
         \left( V_\omega(x) - \E\!\left[V_\omega(0)\right]
         \right) \psi(x)
      \right|^2 \d x
      \stackrel{r\to\infty}{\longrightarrow} 0
   \hfill\widecr}
\ee
for all $\psi\in{\cal S}(\rz^2)$, $\left|\zeta\right|<1$. The last
inequality is due to the estimate
$\Theta\!\left(\left|\zeta-\frac{x}{r}\right|-1\right)\leq
\Theta\!\left(\left|x\right|-r\left(1-\left|\zeta\right|\right)\right)$
and its right-hand side vanishes as $r\to\infty$, because the 
mapping \xref{5b} is almost surely polynomially bounded.

The strong resolvent convergence \xref{A.27} now implies
\cite[Theorem~VIII.20]{resi80} together with \xref{A.21}
\be{A.30}
   \lim_{r\to\infty}
   \E\!\left[\left< \phi_n,\e^{-tE_n\Vregreg E_n}\,\phi_n
   \right>\right]
   =
   \frac{2\pi}{B}\,\e^{t\,\E\left[V_\omega(0)\right]}\, \widetilde R_n(t).
\ee
{}From $\Vregreg(x)\geq -\E\!\left[V_\omega(0)\right]$ one has
\be{A.31}
   \left< \phi_n , \e^{-tE_n\Vregreg E_n}\,\phi_n \right>
   \leq
   \\e^{t\,\E\left[V_\omega(0)\right]}
\ee
almost surely. Therefore one can use the dominated-convergence theorem, 
\xref{A.30} and \xref{A.25} to obtain \xref{A.24}.\halmos
%--------------------------------------------------------------------
\subsubsection*{Proof of \xref{22}}
Let $\Vreg$ be as given in \xref{A.23} and $0<r<\infty$. We first note
that $E_n\Vreg E_n$ is almost surely of trace class. This can be seen
\cite[Theorem~VI.22(h)]{resi80}, for example, by noting that both
$E_n\left|\Vreg\right|^{1/2}$ and ${\rm sgn}(\Vreg)
\left|\Vreg\right|^{1/2}E_n$ are almost surely Hilbert-Schmidt because
\cite[Theorem~VI.23]{resi80}
\be{A.32}
   \int\left| E_n(x,y)\left|\Vreg(y)\right|^{1/2}
   \right|^2 \d x \, \d y
   =
   \frac{B}{2\pi}\int \left|\Vreg(y)\right| \d y
   < \infty.
\ee
Therefore
\be{A.33}
   \e^{-tE_n\Vreg E_n} -1
   =
   E_n\Vreg E_n \:
   \sum_{k=1}^\infty \frac{(-t)^k}{k!} \left(E_n\Vreg E_n\right)^{k-1}
\ee
is of trace class, because it is the product of a trace-class operator
and a norm-convergent sum \cite[Theorem~VI.19(c)]{resi80}. Thus we have
\be{A.34}
   \matrix{\noalign{\vskip\displaystylematrixsep}\Ds &\Ds
      \int
         \E\!\left[\left( E_n \left(\e^{-tE_n\Vreg E_n}-1\right) E_n
         \right)(x,x)\right]
      \d x
   \hfill\widecr\Ds\hfill = &\Ds
      \E\!\left[\tr\!\left\{ E_n \left(\e^{-tE_n\Vreg E_n}-1\right) E_n
      \right\}\right]
   \hfill\widecr}
\ee
due to the continuity of the integral kernel
\cite[Example~X.1.18]{kat84}, see Remark \ref{2-2-iii)}.

Analogously to the arguments in the proof of \xref{21} we may 
rewrite the integrand in the approximation formula \xref{A.24} as a 
scalar product, use the
Jensen-Peierls inequality and in a next step the Jensen inequality
together with $\E\!\left[\Vreg(x)\right]=0$ to show
\be{A.35}
   \E\!\left[\left( E_n \left(\e^{-tE_n\Vreg E_n}-1\right) E_n
   \right)(x,x)\right]
   \geq
   \frac{B}{2\pi}\,
   \E\!\left[\e^{- \frac{2\pi}{B} \, t\,\left(E_n \Vreg E_n\right)(x,x)}-1 \right]
   \geq 0 .
\ee
Thanks to \xref{A.35}, \xref{A.34} and the approximation formula
\xref{A.24} can be combined to yield the preliminary estimate
\be{A.36}
   \e^{t\,\E\left[V_\omega(0)\right]}\, \widetilde R_n(t) -
   \frac{B}{2\pi}
   \leq
   \limsup_{r\to\infty} \frac{1}{\pi r^2}
   \E\!\left[\tr\!\left\{ E_n \left(\e^{-tE_n\Vreg E_n}-1\right) E_n
   \right\}\right].
\ee

Finally, we use the Jensen-Peierls inequality in the version of Berezin 
\cite{ber72}, \cite[Section~8(c)]{sim79b}, which implies
\be{A.37}
   \tr\!\left\{ E_n \left(\e^{-tE_n\Vreg E_n}-1\right) E_n \right\}
   \leq
   \tr\!\left\{ E_n \left(\e^{-t\Vreg}-1\right) E_n \right\}.
\ee
This is justified, since both $E_n\Vreg E_n$ and $\Vreg$ are almost
surely essentially self-adjoint on ${\cal S}(\rz^2)$, $E_n{\cal
S}(\rz^2)$ is dense in $E_n\L^2(\rz^2)$ and because the right-hand side
is well-defined. The latter can be seen by an argument analogous to that
at the beginning of this proof employing
$\left(\e^{-t\Vreg}-1\right)\in\L^1(\rz^2)$. Due to the continuity of
the integral kernel of $E_n \left(\e^{-t\Vreg}-1\right) E_n$ we are
allowed \cite[Example~X.1.18]{kat84} to calculate
\be{A.38}
   \matrix{\noalign{\vskip\displaystylematrixsep}\Ds &\Ds
      \E\!\left[\tr\!\left\{ E_n \left(\e^{-t\Vreg}-1\right) E_n
      \right\}\right]
   \hfill\widecr\Ds\hfill = &\Ds
      \frac{B}{2\pi} \, \E\!\left[
         \int \left(\e^{-t\Vreg(x)}-1\right) \d x
      \right]
   \hfill\widecr\Ds\hfill = &\Ds
      \frac{B r^2}{2} \, \left(
      \e^{t\,\E\left[V_\omega(0)\right]}\,
      \E\!\left[\e^{-t V_\omega(0)}\right] - 1
      \right).
   \hfill\widecr}
\ee
To finish the proof of \xref{22}, we only have to insert \xref{A.37}
into \xref{A.38} and perform the limit with the help of
\xref{A.38}.\halmos
%-----------------------------------------------------------------------------
\end{appendix}
%-----------------------------------------------------------------------------
\section*{Acknowledgments}
We thank L.~A.~Pastur for stimulating comments. 
We are much indebted to W.~Fischer and P.~M\"uller for constructive
criticism.
K.~B.~would like to thank the members of the Mathematics Department of
the Ruhr-Universit\"at Bochum for their kind hospitality, made
possible by financial support of the Sonder\-forschungs\-bereich~237:
\lq\lq Unordnung und gro{\ss}e Fluktuationen\rq\rq.
D.~H.~would like to thank the \lq\lq Graduiertenkolleg Geometrie und
Mathematische Physik an der Ruhr-Universit\"at Bochum\rq\rq\ for
financial support.  Moreover, this work was partially supported by the
Deutsche Forschungsgemeinschaft under Grant No.~Le~330/5-1.
%-----------------------------------------------------------------------------
\raggedright

%-----------------------------------------------------------------------------
\end{document}